\documentclass{ifacconf}

\usepackage{graphicx}      
\usepackage{natbib}        

\usepackage{amsmath,amssymb,amsfonts,bm}
\usepackage{algorithm}
\usepackage{algpseudocode}
\usepackage{graphicx}
\usepackage{tikz,pgfplots}
\usepackage{epsfig} 
\usepackage{pdfpages}
\usepackage{selinput}\SelectInputMappings{adieresis={ä},germandbls={ß}}

\newtheorem{remark}{\bf{Remark}}

\begin{document}
	\begin{frontmatter}
		
		\title{Fast IMU-based Dual Estimation of Human Motion and Kinematic Parameters via Progressive In-Network Computing\thanksref{footnoteinfo}}
		
		\thanks[footnoteinfo]{
		This work is supported by the Federal Ministry of Education and Research of Germany in the programme of “Souverän. Digital. Vernetzt. ” Joint project 6G-life (Project ID: 16KISK002, 16KISK001K), and by the German Research Foundation (DFG) under the grant number 315177489 as part of the SPP 1914 (CPN), and by the European Union’s Horizon 2020 research and innovation programme under the Marie Sk\l{}odowska-Curie grant agreement number 899987, and by the German Research Foundation (DFG, Deutsche Forschungsgemeinschaft) as part of Germany's Excellence Strategy—EXC 2050/1—Cluster of Excellence “Centre for Tactile Internet with Human-in-the-Loop” (CeTI) of Technische Universität Dresden (Project ID: 390696704).
		}
		
		\author[Author_TUM]{Xiaobing Dai} 
		\author[Author_TUDComNets]{Huanzhuo Wu} 
		\author[Author_TUM]{Siyi Wang}
		\author[Author_TUM]{Junjie Jiao}
		\author[Author_TUD]{Giang T. Nguyen}
		\author[Author_TUDComNets]{Frank H. P. Fitzek}
		\author[Author_TUM]{Sandra Hirche}
		
		\address[Author_TUM]{Chair of Information-oriented Control, TU Munich, Germany \\(e-mail: xiaobing.dai, siyi.wang, junjie.jiao, hirche@tum.de)}
		\address[Author_TUDComNets]{Deutsche Telekom Chair of Communication Networks, Centre for Tactile Internet with Human-in-the-Loop (CeTI), TU Dresden, Germany (e-mail: huanzhuo.wu, frank.fitzek@tu-dresden.de)}
		\address[Author_TUD]{Haptic Communication Systems, Centre for Tactile Internet with Human-in-the-Loop (CeTI), TU Dresden, Germany \\(e-mail: giang.nguyen@tu-dresden.de)}
		
		\begin{abstract}                
			Many applications involve humans in the loop, where continuous and accurate human motion monitoring provides valuable information for safe and intuitive human-machine interaction. Portable devices such as inertial measurement units (IMUs) are applicable to monitor human motions, while in practice often limited computational power is available locally. The human motion in task space coordinates requires not only the human joint motion but also the nonlinear coordinate transformation depending on the parameters such as human limb length. In most applications, measuring these kinematics parameters for each individual requires undesirably high effort. Therefore, it is desirable to estimate both, the human motion and kinematic parameters from IMUs. In this work, we propose a novel computational framework for dual estimation in real-time exploiting in-network computational resources. We adopt the concept of field Kalman filtering, where the dual estimation problem is decomposed into a fast state estimation process and a computationally expensive parameter estimation process. In order to further accelerate the convergence, the parameter estimation is progressively computed on multiple networked computational nodes. The superiority of our proposed method is demonstrated by a simulation of a human arm, where the estimation accuracy is shown to converge faster than with conventional approaches.
		\end{abstract}
		
		
	\end{frontmatter}
	
	\section{Introduction}
	\label{sec:introduction}
	
	In many human-machine interaction applications, the accurate estimation of human motion plays an important role.
    In most cases, portable sensors -- in particular inertial measurement units (IMUs) -- are  preferred, due to their flexibility (\cite{vargas2016imu,joukov2017human,yi2021transpose}).

	
    In the literature, most of the methods transform the human motion estimation problem into the orientation estimation problem of IMU sensors.
    In \cite{mahony2005complementary}, a complementary filter is proposed based on the comparison of the measurement with the gravity field and earth magnetic field.
    However, this filter is applicable only to static or quasi-static scenarios.
    In order to deal with dynamic cases, in \cite{alatise2017pose}, an approach based on the extended Kalman filter (EKF) is applied using the raw data from the accelerometer and gyroscope.
	Yet, in this approach, the drift error suffers from the gyroscope, which can be significantly reduced by introducing motion constraints, e.g. human structure.
	To this end, there are works that introduce motion constraints by exploiting the human structure in order to improve the accuracy of human motion estimation from IMU measurement data.
    For example, \cite{lin2012human} consider a serial kinematics model for human with motion constraints described by Denavit–Hartenberg parameters.
    However, the works above only focus on the estimation accuracy of joint motion, ignoring the effect of uncertain human kinematic parameters, e.g., forelimb length, on human motion estimation in task spaces.
    Obtaining accurate human kinematic parameter information in advance for each individual is often infeasible or at least, and reduces usability of the system.
	Motivated by this observation, in the present paper we consider the dual estimation of human motion and the unknown human kinematic parameters, without any additional effort for additional individual kinematic parameter measurements.
	
	In the context of Kalman filter, dual estimation is investigated.
	Based on the vanilla Kalman filter, the constant parameters are treated the same as the hidden state variables but with a lower changing rate, see e.g. \cite{berkane2017attitude,bloesch2013state}.
    However, the changing parameters may deteriorate the state estimation accuracy especially with poor choices of the noise variance for parameters.
    \cite{bania2016field} proposes a field Kalman filter (FKF) from Bayesian view, decomposing the dual estimation problem into computationally fast state estimation and slow parameter estimation.
    In their work, a moving horizon-based approximation method is presented to accelerate the parameter estimation part by using only a sub set of the data, yet this comes at the cost of reduced estimation accuracy.
    Given the ubiquity of communication and computation, we ask how to exploit available distributed computational resources in the network in order to improve the convergence speed of dual estimation.
    Our aim is to facilitate accurate and real-time human motion estimation in the presence of uncertain kinematic parameters.

    The main contribution of the present work is the proposal of a progressive in-network algorithm for dual estimation, providing both the real-time performance of state estimation and fast convergence of parameter estimation.
    In particular, we leverage the progressive principle in \cite{wu2021computing,wu2022picaextension}, and exploit external computational resources to accelerate parameter estimation.
    In our setting, the external computational resources are represented by connected computing nodes \textit{in the network}.
	With \textit{progressively} increasing data set at the subsequent nodes, the accuracy of the parameter estimation is \textit{progressively} improved.
    The performance of our method is illustrated by a numerical evaluation.
	In the evaluation, in-network computation structures with different number of intermediate nodes are simulated for a human arm with 2 degrees of freedom.
	The results show that the proposed progressive in-network dual estimation method significantly reduces the computation time for parameters by $40\%$ without losing any estimation accuracy compared with FKF using the entire data set.
		
	
	The remainder of this paper is structured as follows.
	First, the kinematics model and the output model based on IMUs are introduced in Section \ref{section_modeling}.
	In Section \ref{section_dual_estimation}, the state and parameter dual estimation method is proposed and a progressive in-network algorithm is introduced for the reduction of the computation latency.
	Section \ref{section_evaluation} covers the numerical evaluation and result discussions.
	Finally, Section \ref{sec:conclusion} concludes the contribution.
	
	\section{Preliminaries} \label{section_modeling}
	Before introducing our proposed algorithm for dual estimation, in this section, the human model is discussed.
	In particular, the forward kinematics is introduced in Subsection \ref{subsection_model_kinematics} using the minimal coordinate method.
	Furthermore, the output model based on the measurements from the IMUs is derived in Subsection \ref{subsection_model_IMUs}.
	
	\subsection{Forward Kinematics} \label{subsection_model_kinematics}
	
	\begin{figure}[t]
		\centerline{\includegraphics[width=0.72\columnwidth]{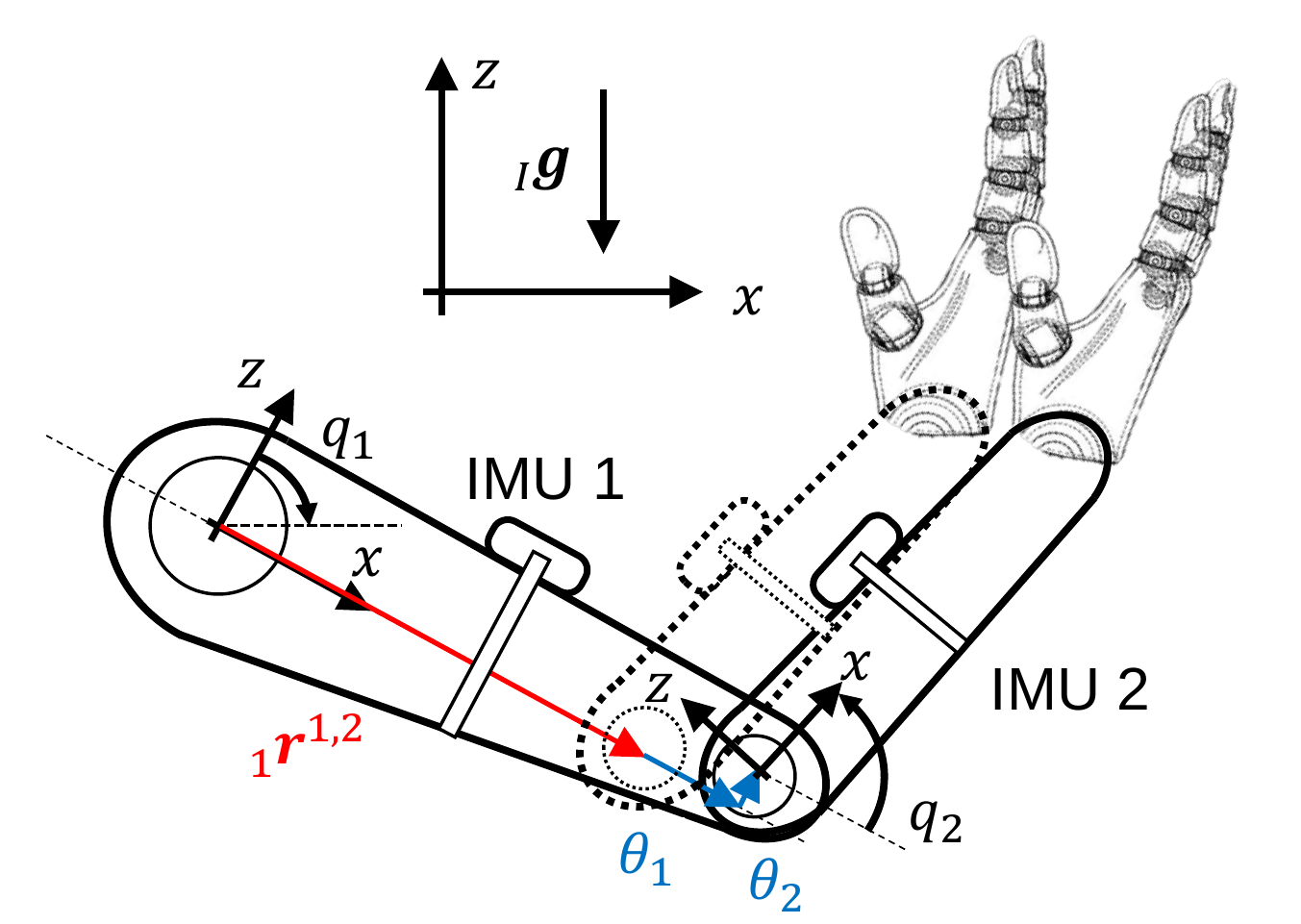}}
		\caption{The human arm model includes planar shoulder and elbow movement, with the generalized coordinate $\bm{q} = [q_1,q_2]^T$, where $q_1$ and $q_2$ are the relative rotation angle of shoulder and elbow along $y$-axis, respectively.
        There are 2 IMUs attached on the upper limb (IMU 1) and forearm (IMU 2). The length ${_1} \bm{r}^{1,2}$
		of the upper limb is described as an unknown parameter $\bm{\theta}$.}
		\label{fig_HumanArm2DoF}
	\end{figure}
	
	For human motion modeling, a serial kinematic chain composed of links connected only by rotational joints along a relatively fixed axis is employed, e.g., the human arm model in Fig. \ref{fig_HumanArm2DoF}.
	The generalized coordinate $\bm{q} \in \mathbb{R}^n$ is chosen as the rotation angle of each joint, where $n \in \mathbb{N}_+$ represents the number of degrees of freedom (DoF).
	The state $\bm{x}_k$ of the system at time $t_k$ is a concatenation of the generalized coordinate $\bm{q}_{k}$, generalized velocity $\dot{\bm{q}}_{k}$ and generalized acceleration $\ddot{\bm{q}}_{k}$, i.e., 
	\begin{align} \label{eqn_system_state}
		\bm{x}_{k} = [\bm{q}^T_{k},\dot{\bm{q}}^T_{k}, ~ \ddot{\bm{q}}^T_{k}]^T \in \mathbb{R}^{3n}.
	\end{align}
	The motion of the links includes both the translational part and the rotational part, and is calculated through forward kinematics.
	
	\subsubsection{1) Forward Rotational Kinematics}

	In the present work, the axis-angle representation as in \cite{gao2021dynamics} is used to express the rotation vector $\bm{\varphi} \in \mathbb{R}^3$ to avoid the singularity.
	The corresponding rotation matrix $\bm{R} \in \mathbb{R}^{3 \times 3}$ is calculated through Rodrigues' formula in \cite{kovacs2012rotation}
	\begin{equation}
		\label{eqn_Rodrigues}
		\bm{R}(\bm{\varphi}) = \bm{I}_3 + \frac{\sin{\theta}}{\theta} \bm{\varphi}_{\times} + \frac{1 - \cos{\theta}}{\theta ^ 2} \bm{\varphi}_{\times} \bm{\varphi}_{\times},
	\end{equation}
	where $\theta = \|\bm{\varphi}\| \in \mathbb{R}$.
    The function $(\cdot)_{\times}: \mathbb{R}^3 \to \mathbb{R}^{3 \times 3}$ is defined as the mapping that associates any vector $\bm{a} = [a_1,a_2,a_3]^T \in \mathbb{R}^3$ with a corresponding skew-symmetric matrix $\bm{a}_{\times} \in \mathbb{R}^{3 \times 3}$, which is expressed as
	\begin{equation*}
		\bm{a}_{\times} = \begin{bmatrix}
			0 & -a_3 & a_2 \\
			a_3 & 0 & -a_1\\
			-a_2 & a_1 & 0
		\end{bmatrix}.
	\end{equation*}
	The Rodrigues' formula for axis-angle representation has only a continuous undefined point with $\bm{R}(\bm{0}_{3 \times 1}) = \bm{I}_3$.
	For a serial kinematics chain, the motion of link $(i+1)$ is only related to the motion of its predecessor $i$ and relative rotation $q_{i+1}, \dot{q}_{i+1}, \ddot{q}_{i+1}$ along the axis $_{i}\bm{n} \in \mathbb{R}^3$, which is fixed in the local coordinate $i, \forall i = 1, \cdots, n$.
	The rotation matrix $\bm{R}_{i+1}$ representing the pose of the $(i+1)$-th link is
	\begin{align} \label{eqn_rotation_matrix_forward}
		\bm{R}_{i+1} = \bm{R}_{i} \bm{R} (_{i}\bm{n} q_{i+1}) = \bm{R}_{i} \bm{R}_{q,i+1}
	\end{align}
	with $\bm{R}_{q,i+1} = \bm{R} (_{i}\bm{n} q_{i+1})$.
	The angular velocity ${_{i} \bm{\omega}^{i}}$ and acceleration ${_{i} \dot{\bm{\omega}}^{i}}$ are
	\begin{align} \label{eqn_forward_rotational_kinematics}
		{_{i+1} \bm{\omega}^{i+1}} = &\bm{R}_{q,i+1}^T {_{i} \bm{\omega}^{i}} + {_{i}\bm{n}} \dot{q}_{i+1}, \nonumber \\
		{_{i+1} \dot{\bm{\omega}}^{i+1}} = &\bm{R}_{q,i+1}^T {_{i} \dot{\bm{\omega}}^{i}} + {_{i+1} \bm{\omega}^{i+1}_{\times}} {_{i}\bm{n}} \dot{q}_{i+1} + {_{i}\bm{n}} \ddot{q}_{i+1}, 
	\end{align}
	where $_{j} c^{i}$ represents the variable $c$ for the $i$-th link in the local coordinate of body $j$.
	The rotational properties for the initial body are set as $\bm{R}_{0}$, ${_{0} \bm{\omega}^{0}}$ and ${_{0} \dot{\bm{\omega}}^{0}}$, which are based on the environment and the motion of the human trunk.
	Note that the rotation axis ${_{i}\bm{n}}$ of human joints, e.g., elbow or wrist, is almost identical even for different people, and thus can be pre-defined and fixed.
	
	\subsubsection{2) Forward Translational Kinematics}
	Since only the rotational freedom is considered, the origin of the $(i+1)$-th link is considered as one fixed point connected on the predecessor link $i$.
	The relative position between the origin of the link $i$ and $i+1$ is given as a constant $_{i} \bm{r} ^{i,i+1}$ defined in the local coordinate $i$.
	The translational motion of the origin of the $(i+1)$-th link then is computed through the motion of link $i$ and $_{i} \bm{r} ^{i,i+1}$ as
	\begin{align} \label{eqn_forward_translational_kinematics}
		&_I \bm{r} ^{i+1} = {_I \bm{r} ^{i}} + \bm{R}_{i} {_{i} \bm{r} ^{i,i+1}}, \nonumber\\
		&_I \dot{\bm{r}} ^{i+1} = _I \dot{\bm{r}} ^{i} + \bm{R}_{i} {_{i} \bm{\omega}^{i}_{\times}} {_{i} \bm{r} ^{i,i+1}},\\
		&_I \ddot{\bm{r}} ^{i+1} = _I \ddot{\bm{r}} ^{i} + \bm{R}_{i} {_{i} \bm{\omega}^{i}_{\times}} {_{i} \bm{\omega}^{i}_{\times}} {_{i} \bm{r} ^{i,i+1}} + \bm{R}_{i} {_{i} \dot{\bm{\omega}}^{i}_{\times}} {_{i} \bm{r} ^{i,i+1}},  \nonumber
	\end{align}
	where the left subscript $I$ represents the inertial coordinate.
	The translational motion of link $i = 0$ is based on the pre-defined constants $ _I\bm{r} ^{0}$, $_I \dot{\bm{r}} ^{0}$ and $_I \ddot{\bm{r}} ^{0}$.
	With the above recursive update method, the translational kinematics properties for each link in the serial chain are obtained.
	
	From the above analysis, a strong relationship is observed between the translational motion and kinematic parameters reflected by $_{i}\bm{r}^{i,i+1}$.
	Considering the different kinematic parameters of each individual, vectors $_{i}\bm{r}^{i,i+1}, \forall i = 1, \cdots, n-1$ are not fixed and thus regarded as unique constant parameters for each human model.
	For the human model, e.g., the human arm, with $N$ bodies, the overall parameters are concatenated as
	\begin{align} \label{eqn_model_parameter}
		\bm{\theta} = [{_{1}\bm{r}^{1,2}}^T, \cdots, {_{N-1}\bm{r}^{N-1,N}}^T]^T \in \mathbb{R}^{3(N-1)},
	\end{align}
	which will also be estimated in Subsection \ref{subsection_dual_estimation}.
	
	\subsection{Measurements from IMU Sensors} 
	\label{subsection_model_IMUs}
	
	In our setting, the motion of the human is reconstructed based on the measurements from IMU sensors.
	The $j$-th IMU is attached to link $i$ with fixed relative position $_i \bm{r} ^{i,j}_s$ and fixed rotation $_i \bm{\varphi} ^{i,j}_s$, where the right subscript $s$ indicates the variable defined for sensors.
	Due to the low reliability of magnetic field measurements because of external disturbances, only the measurements from the gyroscope and the accelerometer are used for motion estimation.
	
	The gyroscope returns the angular velocity $_j\bm{\omega}^j$ in the local coordinate of the IMU $j$, which is expressed as
	\begin{align*} 
		&{_{j} \bm{\omega}^{j}} = \bm{R}_j^T {\bm{R}_i} _i\bm{\omega} ^i + \bm{v}_{\omega,j}, \\
		&\bm{R}_j = \bm{R}_i\bm{R}(_i \bm{\varphi} ^{i,j}_s).
	\end{align*}
	The vector $\bm{v}_{\omega,j}$ is the measurement noise from the gyroscope satisfying independent, identical and zero-mean Gaussian distribution, i.e., $\bm{v}_{\omega,j} \sim \mathcal{N}(\bm{0}_{3\times1}, \bm{Q}_{\omega,j})$ and $\bm{Q}_{\omega,j} = \mathrm{diag}(\sigma_{\omega,1,j}^2,\sigma_{\omega,2,j}^2,\sigma_{\omega,3,j}^2)$.
	The rotation matrix $\bm{R}_i$ and angular velocity $_i\bm{\omega} ^i$ of the link $i$ are iteratively obtained through \eqref{eqn_rotation_matrix_forward} and \eqref{eqn_forward_rotational_kinematics}.
	
	The accelerometer records not only the translational acceleration from the motion, but also the earth gravity ${_I\bm{g}} \in \mathbb{R}^3$.
	Moreover, the recorded acceleration $_j{\bm{a}}^j$ is expressed in the IMU local coordinate $j$ with
	\begin{align*} 
		_j{\bm{a}}^j = \bm{R}_j^T \left(_I \ddot{\bm{r}} ^i + \bm{R}_i (_i \dot{\bm{\omega}} ^i_{\times} + {_i \bm{\omega} ^i_{\times}} {_i \bm{\omega} ^i_{\times}} ) _i \bm{r} ^{i,j}_s + {_I\bm{g}} \right) + \bm{v}_{a,j},
	\end{align*}
	where $\bm{v}_{a,j} \in \mathbb{R}^3$ represents the measurement noise from the accelerometer, which is assumed to follow the independent, identical and zero-mean Gaussian distribution, i.e., $\bm{v}_{a,j} \sim \mathcal{N}(\bm{0}_{3\times1}, \bm{Q}_{a,j})$ with $\bm{Q}_{a,j} = \mathrm{diag}(\sigma_{a,1,j}^2,\sigma_{a,2,j}^2,\sigma_{a,3,j}^2)$.
	The motion properties $_I \ddot{\bm{r}} ^i$ and $_i \dot{\bm{\omega}} ^i$ are obtained from \eqref{eqn_forward_translational_kinematics} and \eqref{eqn_forward_rotational_kinematics}, respectively.
	According to the analysis in Subsection \ref{subsection_model_kinematics}, the translational acceleration $_j{\bm{a}}^j$ is also related to the kinematic parameters $\bm{\theta}$ defined in \eqref{eqn_model_parameter}.
	
	Concatenate the measurements from $M \in \mathbb{N}_+$ IMUs, the output of the system is expressed as $\bm{y} \in \mathbb{R}^{6M}$ with
	\begin{align} \label{eqn_IMU_measurement}
		&\bm{y} = \begin{bmatrix}
			\bm{y}^1 \\ \vdots \\ \bm{y}^M
		\end{bmatrix} = \begin{bmatrix}
			\bm{h}^1(\bm{x},\bm{\theta}) \\ \vdots \\ \bm{h}^M(\bm{x},\bm{\theta})
		\end{bmatrix} + \begin{bmatrix}
			\bm{v}_1 \\ \vdots \\ \bm{v}_M
		\end{bmatrix} = \bm{h}(\bm{x},\bm{\theta}) + \bm{v},
		\\
		&\bm{y}^j = \begin{bmatrix}	{_j{\bm{a}}^j} \\ {j} \bm{\omega}^{j} \end{bmatrix} = \bm{h}^j(\bm{x}, \bm{\theta}) + \bm{v}_{j} \in \mathbb{R}^6, \nonumber
	\end{align}
	where the concatenated measurement noises $\bm{v}_j \sim \mathcal{N}(\bm{0}_{6 \times 1}$, $\bm{Q}_{j}), \forall j = 1,\cdots, M$ and $\bm{v} \sim \mathcal{N}(\bm{0}_{6M \times 1}, \bm{Q}_{v})$.
	The covariance matrix of measurement noise $\bm{Q}_{v} $ is written as $\bm{Q}_{v} = \mathrm{blkdiag}(\bm{Q}_1, \cdots, \bm{Q}_M)\in \mathbb{R}^{6M \times 6M}$, where $\bm{Q}_j = \mathrm{blkdiag}(\bm{Q}_{a,j}, \bm{Q}_{\omega,j}) \in \mathbb{R}^{6 \times 6}$ for all $j = 1,\cdots,M$.
	
	\section{Progressive In-network Dual Estimation}\label{section_dual_estimation}
	
	In this section, our method of progressive in-network state and parameter dual estimation method is introduced.
	In Subsection \ref{subsection_dual_estimation}, the Bayesian principle based dual estimation method with IMUs is first derived.
	Then the progressive in-network algorithm is proposed in Subsection \ref{subsection_progressive_networked_dual_estimation}, to reduce the computation latency from the dual estimation and maintain the estimation accuracy.
	
	\subsection{Dual Estimation of Motion and Kinematic Parameters} \label{subsection_dual_estimation}
	The estimation problem of both motion and kinematic parameters is formulated using the Bayesian principle as the joint probability of states $\bm{x}_k$ in \eqref{eqn_system_state} and parameters $\bm{\theta}$ in \eqref{eqn_model_parameter}, i.e.,
	\begin{align} \label{eqn_Bayesian_principle}
		p(\bm{x}_{k}, \bm{\theta}|\bm{Y}_{k}) = p(\bm{x}_{k}|\bm{Y}_{k}, \bm{\theta}) p(\bm{\theta}|\bm{Y}_{k}),
	\end{align}
	where $\bm{Y}_k = \{\bm{y}_\kappa\}_{\kappa=1}^k$ is the data set containing the previous measurements and available at time $t_k, \forall k \in \mathbb{N}$, in which $\bm{y}_\kappa$ means the measurement \eqref{eqn_IMU_measurement} at time $t_\kappa$.
	From the expression of \eqref{eqn_Bayesian_principle}, the dual estimation problem is decomposed into two parts: state estimation from $p(\bm{x}_{k}|\bm{Y}_{k}, \bm{\theta})$ and parameter estimation from $p(\bm{\theta}|\bm{Y}_{k})$.
	
	\subsubsection{1) Motion Estimation}
	Motion estimation is regarded as the state estimation by solving the posterior distribution $p(\bm{x}_{k}|\bm{Y}_{k}, \bm{\theta})$.
	To obtain $p(\bm{x}_{k}|\bm{Y}_{k}, \bm{\theta})$, the state transmission function is established based on near constant acceleration model (NCAM) in \cite{jazwinski2007stochastic} assuming the constant acceleration of each joint, which is formulated as
	\begin{equation} \label{eqn_StateTransferFunc}
		\bm{x}_{k+1} = \bm{F}_{k} \bm{x}_{k} + \bm{w}_{k}, \quad 
		\bm{F}_{k} = 
		\begin{bmatrix}
			1 & \Delta t_k & \frac{1}{2} \Delta t_k^2  \\
			0 & 1 & \Delta t_k \\
			0 & 0 & 1 
		\end{bmatrix} \otimes \bm{I}_{n},
	\end{equation}
	where the time interval $\Delta t_k$ is defined as $\Delta t_k = t_{k+1} - t_k \in \mathbb{R}_+$, which can be different for each time instances.
	The operator $\otimes$ represents the Kronecker product.
	The 3-rd time derivative of generalized coordinate $\bm{q}$, i.e., the jerk, is considered as the process noise $\bm{w}_{k} \in \mathbb{R}^{3n}$, which is modeled as an independent, identical and zero mean Gaussian distribution, i.e., $\bm{w}_{k} \sim \mathcal{N}(\bm{0}_{3n \times 1},\bm{Q}_w)$.
	By assuming the independence of the jerk in each freedom, $\bm{Q}_w \in \mathbb{R}^{3n \times 3n}$ is written as
	\begin{equation}
		\label{eqn_ProcessErrorCov}
		\bm{Q}_w = 
		\begin{bmatrix}
			\Delta t_k^5/20 & \Delta t_k^4/8  & \Delta t_k^3/6 \\
			\Delta t_k^4/8  & \Delta t_k^3/3  & \Delta t_k^2/2 \\
			\Delta t_k^3/6  & \Delta t_k^2/2  & \Delta t_k
		\end{bmatrix} \otimes \bm{Q}_\varepsilon,
	\end{equation}
	with $\bm{Q}_\varepsilon = \mathrm{diag}(\sigma^2_{\varepsilon,1}, \cdots, \sigma^2_{\varepsilon,n}) \in \mathbb{R}^{n \times n}$.
	Then using the results in \cite{bania2016field}, we get the posterior distribution of the states $\bm{x}_k|\bm{Y}_{k}, \bm{\theta}$ and prior distribution of the measurements $\bm{y}_k|\bm{Y}_{k-1}, \bm{\theta}$, i.e.,
	\begin{align} \label{eqn_KF_result}
		&p(\bm{x}_{k}|\bm{Y}_{k}, \bm{\theta}) = \mathcal{N} (\bm{x}_{k} | \hat{\bm{x}}_{k}(\bm{\theta}), \bm{P}_{k|k}(\bm{\theta})), \\
		&p(\bm{y}_{k}|\bm{Y}_{k-1}, \bm{\theta}) = \mathcal{N} (\bm{y}_{k} | \bm{h}(\bm{F}_{k-1} \hat{\bm{x}}_{k-1},\bm{\theta}), \bm{W}_{k|k-1}(\bm{\theta})), \nonumber
	\end{align}
	with
	\begin{align}
		&\hat{\bm{x}}_{k}(\bm{\theta}) = \bm{F}_{k-1} \hat{\bm{x}}_{k-1}(\bm{\theta}) + \bm{K}_{k}(\bm{\theta}) (\bm{y}_{k} - \bm{h}(\bm{F}_{k-1} \hat{\bm{x}}_{k-1},\bm{\theta})), \nonumber \\
		&\bm{P}_{k|k}(\bm{\theta}) = (\bm{I}_n - \bm{K}_{k}(\bm{\theta}) \bm{H}_{k}(\bm{\theta}) ) \bm{P}_{k|k-1}(\bm{\theta}), \nonumber \\
		&\bm{W}_{k|k-1}(\bm{\theta}) = \bm{H}_{k}(\bm{\theta}) \bm{P}_{k|k-1} \bm{H}_{k}^T(\bm{\theta}) + \bm{Q}_{v}, \nonumber
	\end{align}
	in which
	\begin{align}
		&\bm{K}_{k}(\bm{\theta}) \!=\! \bm{P}_{k|k\!-\!1}(\bm{\theta}) \bm{H}_{k}^T(\bm{\theta}) (\bm{H}_{k}(\bm{\theta}) \bm{P}_{k|k \!-\! 1}(\bm{\theta}) \bm{H}_{k}^T(\bm{\theta}) \!+\! \bm{Q}_v)^{-1} \!, \nonumber \\
		&\bm{P}_{k|k-1}(\bm{\theta}) = \bm{F}_{k-1} \bm{P}_{k-1|k-1}(\bm{\theta}) \bm{F}_{k-1}^T + \bm{Q}_{w}. \nonumber
	\end{align}
	The output of function $\bm{H}_k(\cdot): \mathbb{R}^{3(N-1)} \to \mathbb{R}^{6M \times 3n}$ is defined as the Jacobian matrix between the output $\bm{y}_k$ and the system states $\bm{x}_k$ at time $t_k, \forall k \in \mathbb{N}$, whose detailed expression is shown in Appendix \ref{Appendix_Jacobian}.

	The initial values of the estimated states $\hat{\bm{x}}_{0}$ are usually chosen as the initial states $\bm{x}_0$.
	Under the assumption of known kinematic parameters $\bm{\theta}$, the computation for state estimation in \eqref{eqn_KF_result} has a low computation load with the complexity of $\mathcal{O}(1)$.
	The estimated state uses the mode of $p(\bm{x}_{k}|\bm{Y}_{k}, \bm{\theta})$, i.e., $\hat{\bm{x}}_{k}(\bm{\theta})$, while the prior distribution of the output, i.e., $p(\bm{y}_{k}|\bm{Y}_{k-1}, \bm{\theta})$, will be used for kinematic parameter estimation later.
	
	\subsubsection{2) Kinematic Parameter Estimation}
	Kinematic parameter estimation is conducted through obtaining the distribution of parameters $\bm{\theta}$ by given previous measurements $\bm{Y}_k$, i.e., $\bm{\theta}|\bm{Y}_{k}$, whose probability is written as
	\begin{align} \label{eqn_parameter_estimation}
		p(\bm{\theta}|\bm{Y}_{k}) = \frac{p(\bm{\theta})}{p(\bm{Y}_{k})}  \prod_{\kappa=1}^{k} p(\bm{y}_{\kappa}|\bm{Y}_{\kappa-1}, \bm{\theta}),
	\end{align}
	where $p(\bm{\theta})$ is the prior distribution of the parameter $\bm{\theta}$, which is manually chosen as $p(\bm{\theta}) = \mathcal{N} (\bm{\theta}_0, \bm{\Sigma}_{\theta,0})$.
	The probability $p(\bm{y}_{\kappa}|\bm{Y}_{\kappa-1}, \bm{\theta})$ is the prior distribution of the measurement, and is calculated from \eqref{eqn_KF_result}.
	However, \eqref{eqn_parameter_estimation} is hard to be calculated due to the multiplication of the Gaussian distributions.
	Hence, a maximum a posteriori probability (MAP) is used to obtain the mode of the distribution in \eqref{eqn_parameter_estimation} as the estimated parameter $\bm{\theta}$.
	MAP converts a problem of obtaining the distribution to an optimization problem for getting the mode of the distribution,
	\begin{align}
		\mathrm{mode}(p(\bm{\theta}|\bm{Y}_{k})) &= \arg\max_{\bm{\theta}} p(\bm{\theta}|\bm{Y}_{k}) = \arg\min_{\bm{\theta}} S_k(\bm{\theta}),
	\end{align}
	where the scalar function $S_k(\bm{\theta}): \mathbb{R}^{3(N-1)} \to \mathbb{R}$ is from negative log-likelihood, i.e., $-\log p(\bm{\theta}|\bm{Y}_{k})$, without some constant terms and scaling irrelevant to the optimization.
	The function $S_k(\bm{\theta})$ is then written as
	\begin{align} \label{eqn_Lk}
		S_k(\bm{\theta}) =& \log |\bm{\Sigma}_{\theta,0}| + (\bm{\theta} - \bm{\theta}_0)^T \bm{\Sigma}_{\theta,0}^{-1} (\bm{\theta} - \bm{\theta}_0)  \\
		& + \sum_{\kappa=1}^{k} \left( \log |\bm{W}_{\kappa|\kappa-1}| + \Delta \bm{y}_{\kappa}^T \bm{W}_{\kappa|\kappa-1}^{-1} \Delta \bm{y}_{\kappa} \right), \nonumber
		\\
		\Delta \bm{y}_{\kappa} = \bm{y}_{\kappa}& - \bm{h}(\bm{F}_{\kappa-1} \hat{\bm{x}}_{\kappa-1},\bm{\theta}), \forall \kappa = 1,\cdots,k . \nonumber
	\end{align}
	Note that the evaluation of $S_k(\bm{\theta})$ requires $\hat{\bm{x}}_{\kappa}$ with $\kappa = 1, \cdots, k$, which are related to the initial states $\bm{x}_0$ and the parameters $\bm{\theta}$.
	This means all the estimated states should be updated when the parameters change, i.e., $k$ times of state estimation in \eqref{eqn_KF_result}.
	Hence, the computation complexity of $S_k(\bm{\theta})$ is $\mathcal{O}(k)$, which enlarges with the growing data set $\bm{Y}_k$ w.r.t time.
	For some nonlinear optimizers, $S_k$ is evaluated many times, which causes a larger computational delay.
	
	Note that, the motion (state) estimation should be in real-time due to the safety consideration in human-machine interaction, which means it always occupies limited local computation resources.
	This blocks the implementation of parameter estimation on the same computation node.
	In our algorithm, the time-consuming parameter estimation is distributed into a serial network for a progressive improvement of its estimation accuracy, and conducted parallel to the state estimation.
	The parameter estimation accuracy is progressively improved through these computation nodes.
	This progressive in-network algorithm for dual estimation is introduced and discussed in the next subsection. \looseness=-1
	
	\subsection{Progressive In-network Dual Estimation}
	\label{subsection_progressive_networked_dual_estimation}
	
\begin{figure}[t]
		\centerline{\includegraphics[width=1\columnwidth]{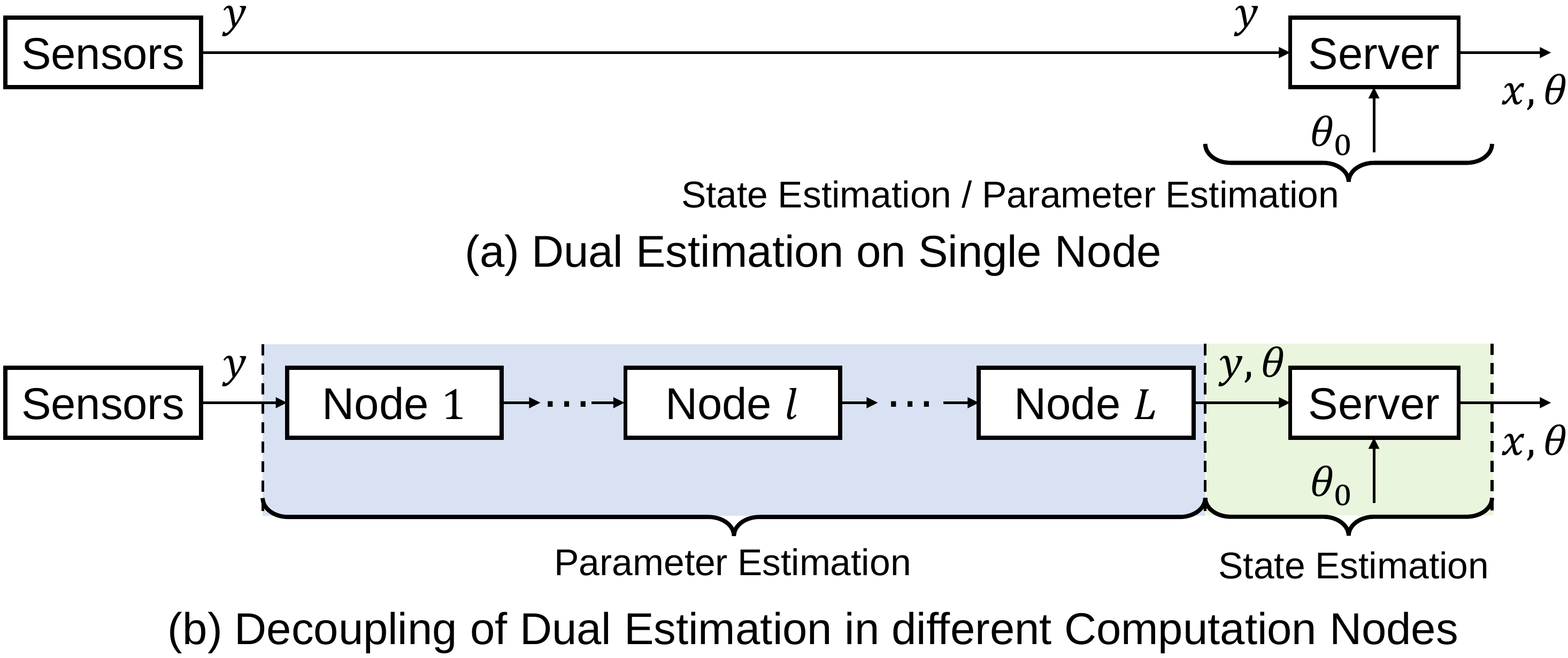}}
		\caption{(a) The dual estimation is implemented only on one node, i.e., the server; (b) The dual estimation is decomposed and distributed to different nodes. Nodes $1, \cdots, L$ are the intermediate computation nodes for parameter estimation. The newly updated parameters $\bm{\theta}$ from each intermediate node are transmitted to its successors and eventually reach the server. The server implements the state estimation using currently available parameters.
		Note that (a) can be regarded as a special case of (b) with $L = 0$.}
		\label{fig_Structure}
	\end{figure}

	To reduce the computation latency and keep the estimation accuracy by using the full data set, the decomposed dual estimation is distributed in multiple computation nodes with the structure in Fig. \ref{fig_Structure}(b).

    \subsubsection{1) Design Principle}
	We consider a structure shown in Fig \ref{fig_Structure}(b) with totally $(L+1)$ nodes and $L \in \mathbb{N}_+$.
	The decomposed state estimation and parameter estimation are distributed into different nodes \textit{in the network}.
	The last node $(L+1)$, also called the server, mainly implements the state estimation immediately receiving new measurement data using the parameters from its predecessor nodes.
	Since the state estimation requires few calculations, the server provides the current estimated state almost in real time.
	Each intermediate node $l = 1, \cdots, L$ consists of a router and a virtual network function (VNF).
	The router is mainly used for data transmission of both measurements $\bm{y}$ and parameters $\bm{\theta}$, while the VNF is mainly used for the parameter estimation by minimizing \eqref{eqn_Lk}.
	In the each node $l = 1, \!\cdots\!, (L\!-\!1)$, only a subset of measurements $\bm{Y}_{K_l} \!=\! \{ \bm{y}_{\kappa}\}_{\kappa \!=\! 1}^{K_l}$ is used for parameter estimation.
	The entire data set is used in node $L$ for final parameter estimation.
	
	To accelerate the parameter estimation through the network, the growing strategy and greedy strategy are introduced.
	The growing strategy progressively increases the size of data subset at subsequent nodes in the network for a better estimation accuracy of the parameters.
    The greedy strategy allows the parameter estimation to exit earlier when the marginal gain is too small in one node.
    Hence, our algorithm is named as \textit{progressive in-network dual estimation} and shown in Algorithm \ref{algorithm_intermediate_node}.
	
	\begin{algorithm}
		\caption{Algorithm on node $l$}
		\label{algorithm_intermediate_node}
		\begin{algorithmic} [1]
			\Statex $\mathbf{Input:}$  $\beta_l$, $\bm{Y}_{K_l}$, $\bm{\theta}_{l-1}^*$
			\If{$K_l < \beta_l$} 
				\State {Wait}
			\Else
				\State $\bm{\theta}_{l,0} \gets \bm{\theta}_{l-1}^*$
				\State $p \gets 0$
				\While{True}
					\State $p \gets p + 1$
					\State $\bm{\theta}_{l,p+1} \gets $ \eqref{eqn_GradientDescent} based on $\bm{\theta}_{l,p}$, $\bm{Y}_{K_l}$
					\If{$o^{\Delta}_{l,p} \le \underline{o}^{\Delta}$ \textbf{OR} $o^{\Delta}_{l,p} \le \underline{o}^{\Delta}$} 
						\State $\beta_{l+1} \gets K_l + {\alpha}$
						\State $\bm{\theta}_l^* \gets \bm{\theta}_{l,p+1}$
						\State return
					\EndIf
				\EndWhile
			\EndIf
		\end{algorithmic}
	\end{algorithm}

    To evaluate the marginal gain in each optimization iteration, the optimizer will be discussed first.
	Then the greed strategy and the growing strategy are introduced in detail.
	
	\subsubsection{2) Optimization Solver}
	To evaluate the marginal gain in each optimization iteration, the optimizer based on a gradient descent algorithm for \eqref{eqn_Lk} is used on the node $l$ with the initial guess $\bm{\theta}_{l,0}$.
	Defining $\bm{\theta}_{l,p}$ as the result of parameters in node $l$ after $p$ optimization iterations, the result after the next iteration $p+1$ is calculated through
	\begin{equation} \label{eqn_GradientDescent}
		\bm{\theta}_{l,p+1} = \bm{\theta}_{l,p} - \gamma \Delta \bm{\theta}_{l,p}, \quad \gamma = \lambda / K_l,
	\end{equation}
	where $\lambda$ is a fixed constant and the learning rate $\gamma$ is varying with the size of the data set $K_l$.
	Parameter estimation with a smaller data set may not be accurate, and therefore the parameters can be quickly adjusted in a coarse range with a higher learning rate.
	Estimation with a larger data set produces a more accurate result with a lower learning rate but a slower convergence speed.
	Considering that nodes closer to the server have a larger data set, a gradually decreasing learning rate is reasonable and computationally efficient.

	The vector $\Delta \bm{\theta}_{l,p - 1}$ is the approximation of the gradient $\mathrm{d} S_{K_l}(\bm{\theta}) / \mathrm{d} \bm{\theta}$ at $\bm{\theta}_{l,p}$. 
	For low computational complexity, the approximated gradient $\Delta \bm{\theta}_{l,p - 1}$ is obtained via single-side finite differential approximation, which is written as
	\begin{equation*}
		\Delta \bm{\theta}_{l,p} = \frac{1}{\epsilon} \begin{bmatrix}
			S_{K_l}(\bm{\theta}_{l,p} + \epsilon \bm{s}_{1}) - S_{K_l}(\bm{\theta}_{l,p})\\
			\vdots \\
			S_{K_l}(\bm{\theta}_{l,p} + \epsilon \bm{s}_{3(N-1)}) - S_{K_l}(\bm{\theta}_{l,p})
		\end{bmatrix},
	\end{equation*}
	where $\bm{s}_{i}, \forall i = 1, \cdots, 3(N-1)$ is the $i$-th column of the identical matrix $\bm{I}_{3(N-1)}$ and $\epsilon \in \mathbb{R}$ is a predefined small value.
	Then, the computation complexity of each optimization iteration is fixed by $\mathcal{O}((3N-2){K_l})$.
	
	The iteration times of \eqref{eqn_GradientDescent} on node $l$ depend on the design of the greedy strategy, which is discussed later.
	
	\subsubsection{3) Greedy Strategy}
	Since the intermediate node $l$ has only a subset $\bm{Y}_{K_l}$ of the whole measurement data, the intermediate result from node $l$ is not necessary final and will be optimized with increasing data at next node.
	Hence, the termination criteria are designed to indicate whether the local marginal gain of an iteration round becomes too small.
	The local marginal gain is evaluated through the parameters' change using the infinity norm $o_{l,p}$ and the approximated gradient using the Euclidean norm $o^{\Delta}_{l,p}$, i.e.,
	\begin{equation*}
		o_{l,p} = \| \bm{\theta}_{l,p+1} - \bm{\theta}_{l,p} \|_{\infty}, \quad
		o^{\Delta}_{l,p} = \| \Delta \bm{\theta}_{l,p} \|.
	\end{equation*}
	If $o_{l,p}$ is smaller than a predefined error bound $\underline{o}$, i.e., $o_{l,p} \le \underline{o}$, then the parameter $\bm{\theta}_{l,p+1}$ is accurate enough and further iterations are unnecessary.
	If the current $o^{\Delta}_{l,p}$ appears too small, i.e., $o^{\Delta}_{l,p} \le \underline{o}^{\Delta}$, further iterations will not improve the $\bm{\theta}_{l,p+1}$ significantly anymore.
	Parameter estimation terminates when one of the above criteria is met, i.e., $o_{l,p} \le \underline{o}$ or $o^{\Delta}_{l,p} \le \underline{o}^{\Delta}$.
	The last parameters after satisfying one of the above termination criteria are regarded as the quasi-optimal parameters $\bm{\theta}_l^*$ from node $l$. 
	The parameters $\bm{\theta}_l^*$ are used as the initial guess for the next node, i.e., $\bm{\theta}_{l,0} = \bm{\theta}_{l-1}^*, \forall l = 1, \cdots, L$ with $\bm{\theta}_{0}^* = \bm{\theta}_{0}$.
	Note that the server also receives the intermediate result $\bm{\theta}_l^*$ and conducts the state estimation based on these updated parameters.
	
	Parameter estimation on the $L$-th node only exists when $o_{L,p} \le \underline{o}$, which guarantee the final accuracy of the parameter estimation.
	
	\subsubsection{4) Growing Strategy}
	Smaller input data sets provide less information to obtain the optimized parameters from \eqref{eqn_Lk}, leading to a larger parameter estimation error.
	Therefore, a progressively growing data set is necessary at subsequent nodes.
	It is easy to see, that node closer to the server uses a larger data set, i.e., if $l_2 > l_1$ then $K_{l_2} > K_{l_1}$, because the node receives more data when its predecessor implement the parameter estimation.
	However, the size of the data subset may vary slightly between two nodes, especially if the parameter estimation is completed quickly in the previous node.
    However, it is desirable to have enough new measurements to start a new optimization on the next node.
	Motivated by the above, a dynamical adjustment of the necessary data set size $\beta_l$ is proposed to ensure that enough new data is added to the next node.
	To quantify the change in the data size, a controlling parameter $\alpha_{l}$ is introduced.
	Specifically, the necessary data set size $\beta_{l+1}$ at the next node becomes
	\begin{equation*}
		\beta_{l+1} = K_{l} + \alpha_{l}.
	\end{equation*}
	The proper choice of $\alpha_{l}$ depends on the distribution of input data and the convergence rate of the centralized iteration algorithm. 
	In this way, the size of input data progressively increased to provide more information for parameter estimation to achieve higher accuracy.
	Note that the optimization starts on the $l$-th node only when the collected data set reaches the required minimum size, i.e., $K_l \ge \beta_l$.
	The last node for parameter estimation, i.e., node $L$, always uses the entire data set.
	Since most of the jobs for optimizing $\bm{\theta}$ have been done on previous nodes, little additional effort is required in node $L$. 
	
	\begin{remark}
		The reduction in computation latency by using the progressive in-network dual estimation arises from two aspects.
		First, the parameter estimation starts before the transmission of the whole set of measured data, so the overlap of data transmission time and parameter estimation time is the time we save.
		Second, we first perform the parameter estimation with a smaller data set and use the intermediate results for the next node with a larger data set.
		Additional with the greedy strategy, this avoids the enormous amount of computation on the server caused by directly using the entire data set and initial values far from the optimal value.
		Moreover, since the entire data set is eventually used, the estimation accuracy is not reduced by our algorithm.
	\end{remark}
	
	
	\section{Evaluation}
	\label{section_evaluation}
	
	In this section, a numerical evaluation with shoulder-elbow movement is performed to illustrate the performance of progressive in-network dual estimation.
	The detailed evaluation setup is given in Subsection \ref{sec:evaluation_setup}, and the convergence time and estimation accuracy using the proposed algorithm are analyzed in Subsection \ref{sec:service_time} and Subsection \ref{sec:progressive_param_esti}, respectively.
	
	
	\subsection{Evaluation Setup} \label{sec:evaluation_setup}
		
	In the numerical evaluation, the planar movement of the human arm with shoulder and elbow movement is considered, which has 2 degrees of freedom, i.e., $q_1$ represents the relative rotation of the upper arm and $q_2$ of the forearm in Fig. \ref{fig_HumanArm2DoF}.
	An IMU sensor is attached to each link, for a total of 2 IMUs.
	Due to the evaluation setting, only the kinematic parameters for the link between the shoulder and elbow play a role, i.e., the parameters for upper arm.
	The kinematic parameter of the upper arm is divided into the nominal part $_1\tilde{\bm{r}}^{1,2}$ (same for all people) and the individual part $\bm{\theta}$, i.e., $_1{\bm{r}}^{1,2} = {_1\tilde{\bm{r}}^{1,2}} + \bm{\theta}$. 
    The detailed information is shown in Table \ref{table:experiment_setting}.
    The units in this section are based on International System of Units (SI).
	
	\begin{table}[t]
		\caption{Evaluation Setting}
		\begin{center}
			\begin{tabular}{c|c|c|c|c|c}
				n & N & M & $\Delta t_k, \forall k$ & $\varepsilon$ & $\lambda$\\
				\hline
				$2$ & $2$ & $2$ & $0.01$ & $10^{-6}$ & $10^{-4}$\\
				\hline \hline
				$\sigma^2_{a,i,j}$ & $\sigma^2_{\omega,i,j}$ & $\sigma^2_{\varepsilon,i}$ & ${\bm{\theta}}_{true}$ & $\bm{\theta}_0$ & ${\bm{\Sigma}}_0$  \\
				\hline
				$0.005$ & $0.002^{\circ}$ & $0.5$ & $[0.05;0;0.03]$ & $[0;0;0]$ & $\bm{I}_3$  \\
				\hline \hline
				$\bm{q}_0$ & $\dot{\bm{q}}_0$ & $\ddot{\bm{q}}_0$ & $\bm{q}_e$ & $\dot{\bm{q}}_e$ & $\ddot{\bm{q}}_e$\\
				\hline
				$[0;0]$ & $[0;0]$ & $[0;0]$ & $[\pi/4;\pi/2]$ & $[0;0]$ & $[0;0]$ \\
				\hline \hline
				$_0{\bm{r}}^{0,1}$ & $_1\tilde{\bm{r}}^{1,2}$ & $_0\bm{n}$ & $_1\bm{n}$ & $_I\bm{r}^0$ & $\bm{R}_0$\\
				\hline
				$[0;0;0]$ & $[0.2;0;0]$ & $[0;1;0]$ & $[0;1;0]$ & $[0;0;0]$ & $\bm{I}_3$ \\
				\hline \hline
				$_1 \bm{\varphi} ^{1,1}_s$ & $_2 \bm{\varphi} ^{2,2}_s$ & \multicolumn{2}{c|}{$_1{\bm{r}}^{1,1}_s$}  & \multicolumn{2}{c}{$_2{\bm{r}}^{2,2}_s$}   \\
				\hline
				$[0;\pi;0]$ & $[0;\pi;0]$ & \multicolumn{2}{c|}{$[0.1;0;0.05]$} & \multicolumn{2}{c}{$[0.1;0;0.05]$}
			\end{tabular}
		\end{center}
		\label{table:experiment_setting}
	\end{table}
	
	The designed trajectory in time period $[0,t_e]$ with $t_e=1$ of the joints angle $\bm{q}$ follows a fifth order polynomial function, where $\bm{q}(t=0)=\bm{q}_0$ and $\bm{q}(t=t_e)=\bm{q}_e$ and the first order and second order of time derivative are $\bm{0}$.
	After $t_e$, the human arm remains in its final pose with zero velocity and acceleration.
	The data set is generated as the concatenation of the IMU measurements at each time.
	
	
	In our simulation, evaluations with a different number of intermediate nodes $L$ are considered, $L = 1, \cdots,6$.
	The error bounds for the greedy strategy are set as $\underline{o} = 2\times10^{-4}$ and $\underline{o}^{\Delta} = 30$.
	For each structure with $L$ intermediate nodes, the simulation runs $50$ times to exhibit the statistical characteristic of the performance.
	For comparison, the dual estimation without the in-network computation proposed in \cite{bania2016field} is chosen with no intermediate node, i.e.,  $L = 0$.
	For each setup, the convergence time $T_{\underline{o}}$ for given $\underline{o}$ and the error of parameters $e_{\theta}$ are analyzed, which are defined later. 
	
	The proposed algorithm is implemented in Python, which can be directly deployed as microservices later.
	The simulation is conducted on a Commercial off-the-shelf (COTS) server with an M1 Pro CPU with 16GB RAM using macOS Monterey version 12.6.
	
	
	\subsection{Convergence Time for Dual Estimation}\label{sec:service_time}
	
\begin{figure}[t]
		\centerline{\includegraphics[width=1\columnwidth]{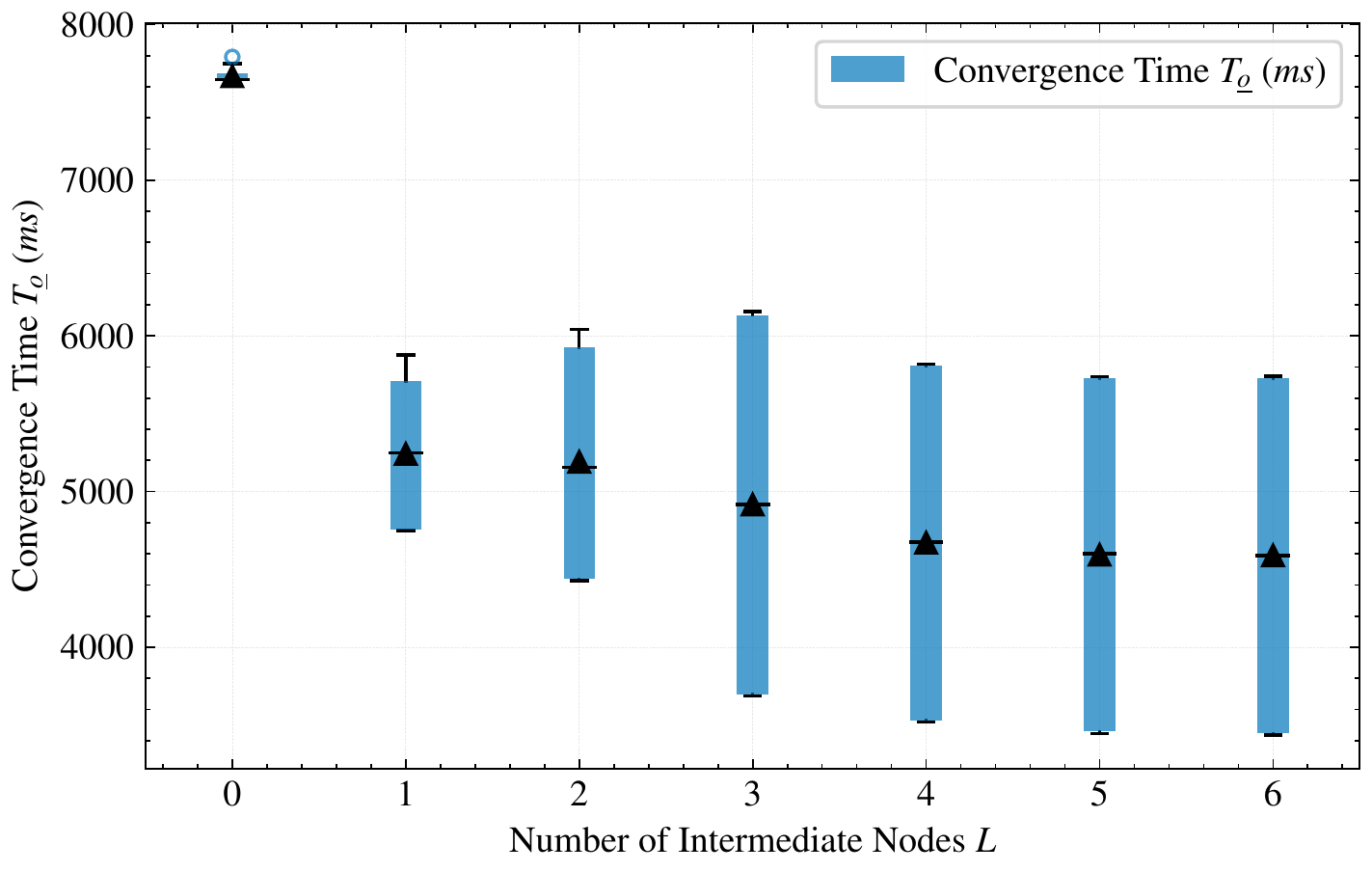}}
		\caption{This figure shows the convergence time of our algorithm with $L$ intermediate nodes. The mean value of the convergence time is marked as a triangle. The bottom and top box edges indicate the 25th and 75th percentiles, respectively. In comparison, $L = 0$ indicates the original method without any intermediate node.
		}
		\label{fig_total_time}
	\end{figure}

	In this subsection, the convergence time $T_{\underline{o}}$ for the proposed progressive in-network dual estimation with different intermediate nodes is compared with the original method with $L = 0$.
	The convergence time $T_{\underline{o}}$ is counted from the generation of the first measurement data to the end of whole dual estimation, i.e., the parameters' change $o_{L,p}$ converge into $\underline{o}$ ($o_{L+1,p} \le \underline{o}$).
	The result is shown in Fig. \ref{fig_total_time}.
	
	First, it is observed using our algorithm $T_{\underline{o}}$ decreases significantly compared to the original method with $L = 0$.
	Besides, $T_{\underline{o}}$ gradually decreases from $7666~ms$ to $4588~ms$ as the number of intermediate nodes increases, i.e., $L$ from $0$ to $6$, which is a $40\%$ acceleration of dual estimation.
	Moreover, the simulation results also show that the acceleration efficiency exhibits a limit as $L$ increases. 
	In this case, adding more intermediate nodes can only improve $T_{\underline{o}}$ little and thus economically inefficient, by considering the additional investment on the hardware.
	
	
	\subsection{Parameter Estimation Error}\label{sec:progressive_param_esti}
	
	Besides the convergence time, the accuracy trends of the intermediate results are also important, since the intermediate results will also be used to improve the accuracy of motion estimation.
	In this subsection, the parameter estimation error defined as $e_{\theta} = \| \bm{\theta} - \bm{\theta}_{L}^* \|$ is analyzed with different $L$.
	
	\begin{figure}[t]
		\centerline{\includegraphics[width=1\columnwidth]{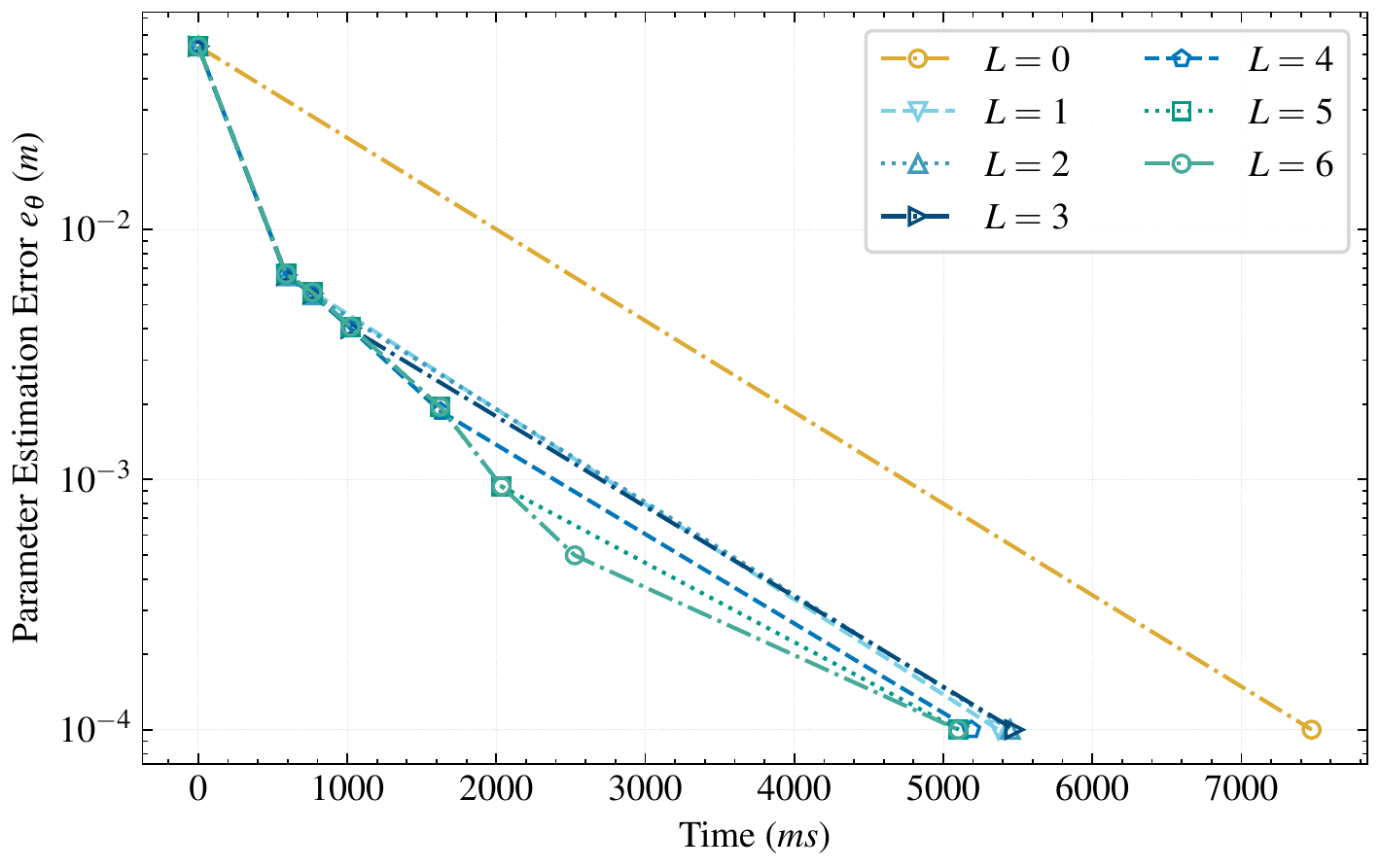}}
		\caption{Accuracy of the intermediate parameters with different intermediate nodes $L$ representing by the parameter error $e_{\theta}$. Besides, it also shows the time, when the intermediate parameters $\bm{\theta}$ are available for the server. Line with $L = 0$ shows no intermediate results with original method.
		}
		\label{fig:optimization_theta}
	\end{figure}
	
	Fig. \ref{fig:optimization_theta} shows the optimization process of the parameters $\theta$ with different number of intermediate nodes $L$.
	For our algorithm $L > 0$, the estimation error $e_{\theta}$ decreases from $0.53$ to $0.08$ at about $800~ms$ in the earliest.
	With terminal criterion and progressively increasing sub data set, the error $e_{\theta}$ continues to decrease with high speed.
	And with more intermediate nodes, more accurate parameters $\bm{\theta}$ are available for the state estimation in the server.
	This evaluation confirms the convergence speed in terms of available time of intermediate results by using our progressive in-network algorithm.
	
	\begin{figure}[t]
		\centerline{\includegraphics[width=1\columnwidth]{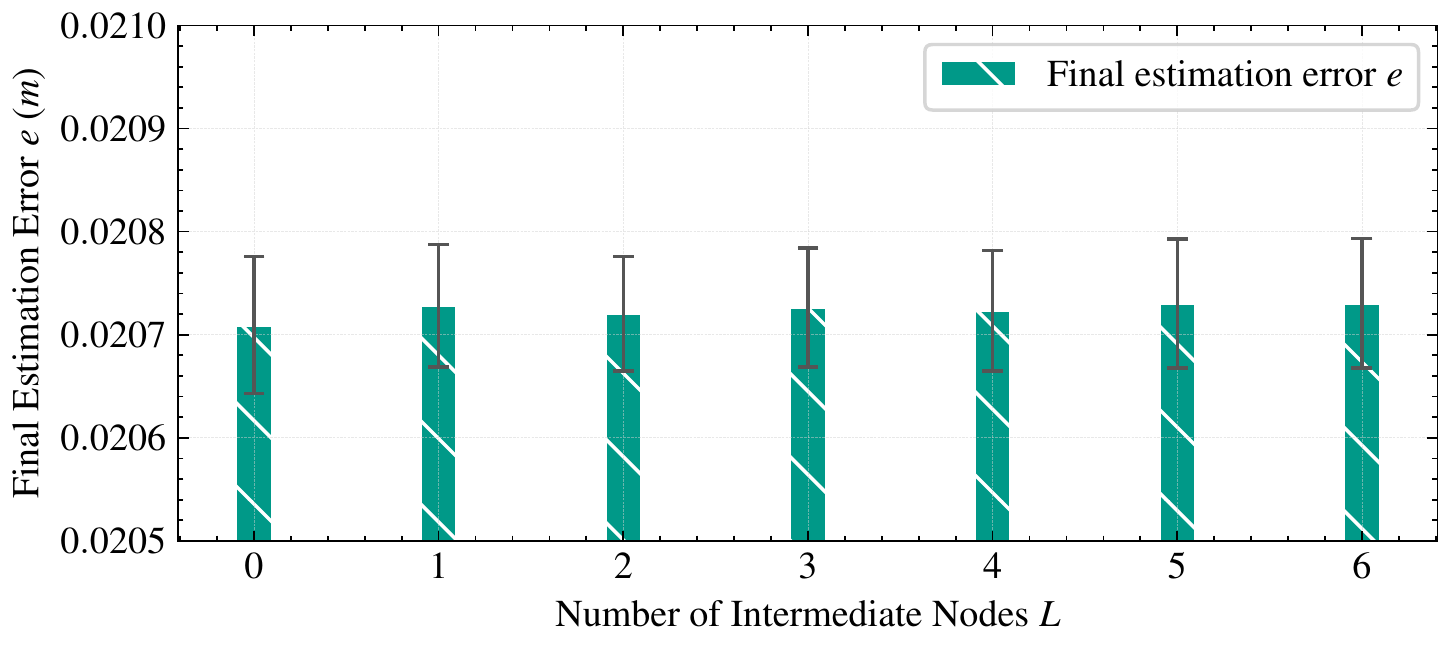}}
		\caption{Final parameter estimation error with $95\%$ confidence interval. The final estimation error of parameters by using our algorithm with different intermediate nodes $L>0$ is similar to the original method with $L = 0$, yet with much faster convergence.
		}
		\label{fig:overall_error}
	\end{figure}
	
	Fig. \ref{fig:overall_error} also shows the final estimation error $e = \| \bm{\theta}_L^* - \bm{\theta}_{true}\|$ for the parameter with different number of intermediate nodes $L$.
	It turns out that there are no significant differences between the final estimation errors for different for different $L = 0,\cdots,6$, i.e., $e \approx 0.02$ comes from an insufficiently large overall data set.
	This finding suggests that the proposed progressive in-network algorithm does not introduce extra estimation error while accelerating the parameter estimation.
	This is a result of the fact that the final termination condition of the proposed method does not change in any way that ensures all data are used for the final iteration, thus guaranteeing the accuracy of the parameter estimation.
	
	Therefore, all effects discussed in the previous sections are observed, which demonstrate the effectiveness of the proposed method.
	
	
	\section{Conclusion} \label{sec:conclusion}
	
	In this paper, we develop the progressive in-network algorithm for dual estimation of human motion and uncertain kinematic parameters simultaneously based on the IMU sensors.
	By distributing the decoupled dual estimation into different nodes, the algorithm guarantees the real-time performance of the human motion estimation and fast convergence of the kinematic parameter estimation, which is vital for human safety in human-machine interaction.
	In the simulation, the convergence time is significantly reduced by using our algorithm without any loss of estimation accuracy.
	
	\bibliography{ifacconf}             

\begin{thebibliography}{14}
\providecommand{\natexlab}[1]{#1}
\providecommand{\url}[1]{\texttt{#1}}
\providecommand{\urlprefix}{URL }
\expandafter\ifx\csname urlstyle\endcsname\relax
  \providecommand{\doi}[1]{doi:\discretionary{}{}{}#1}\else
  \providecommand{\doi}{doi:\discretionary{}{}{}\begingroup
  \urlstyle{rm}\Url}\fi

\bibitem[{Alatise and Hancke(2017)}]{alatise2017pose}
Alatise, M.B. and Hancke, G.P. (2017).
\newblock {Pose estimation of a mobile robot based on fusion of IMU data and
  vision data using an extended Kalman filter}.
\newblock \emph{{Sensors}}, 17(10), 2164.

\bibitem[{Bania and Baranowski(2016)}]{bania2016field}
Bania, P. and Baranowski, J. (2016).
\newblock {Field Kalman filter and its approximation}.
\newblock In \emph{{2016 IEEE 55th Conference on Decision and Control (CDC)}},
  2875--2880. IEEE.

\bibitem[{Berkane and Tayebi(2017)}]{berkane2017attitude}
Berkane, S. and Tayebi, A. (2017).
\newblock {Attitude and gyro bias estimation using GPS and IMU measurements}.
\newblock In \emph{{2017 IEEE 56th Annual Conference on Decision and Control
  (CDC)}}, 2402--2407. IEEE.

\bibitem[{Bloesch et~al.(2013)Bloesch, Hutter, Hoepflinger, Leutenegger,
  Gehring, Remy, and Siegwart}]{bloesch2013state}
Bloesch, M., Hutter, M., Hoepflinger, M.A., Leutenegger, S., Gehring, C., Remy,
  C.D., and Siegwart, R. (2013).
\newblock {State estimation for legged robots-consistent fusion of leg
  kinematics and IMU}.
\newblock \emph{{Robotics}}, 17, 17--24.

\bibitem[{Gao et~al.(2021)Gao, Dai, Kleeberger, and Fottner}]{gao2021dynamics}
Gao, L., Dai, X., Kleeberger, M., and Fottner, J. (2021).
\newblock {Dynamics Modelling and Simulation of Super Truss Element based on
  Non-linear Beam Element}.
\newblock In \emph{{SIMULTECH}}, 50--61.

\bibitem[{Jazwinski(2007)}]{jazwinski2007stochastic}
Jazwinski, A.H. (2007).
\newblock \emph{{Stochastic processes and filtering theory}}.
\newblock {Courier Corporation}.

\bibitem[{Joukov et~al.(2017)Joukov, {\'C}esi{\'c}, Westermann, Markovi{\'c},
  Kuli{\'c}, and Petrovi{\'c}}]{joukov2017human}
Joukov, V., {\'C}esi{\'c}, J., Westermann, K., Markovi{\'c}, I., Kuli{\'c}, D.,
  and Petrovi{\'c}, I. (2017).
\newblock {Human motion estimation on Lie groups using IMU measurements}.
\newblock In \emph{{2017 IEEE/RSJ International Conference on Intelligent
  Robots and Systems (IROS)}}, 1965--1972. IEEE.

\bibitem[{Kov{\'a}cs(2012)}]{kovacs2012rotation}
Kov{\'a}cs, E. (2012).
\newblock Rotation about an arbitrary axis and reflection through an arbitrary
  plane.
\newblock In \emph{Annales Mathematicae et Informaticae}, volume~40, 175--186.

\bibitem[{Lin and Kuli{\'c}(2012)}]{lin2012human}
Lin, J.F. and Kuli{\'c}, D. (2012).
\newblock {Human pose recovery using wireless inertial measurement units}.
\newblock \emph{{Physiological Measurement}}, 33(12), 2099--2115.

\bibitem[{Mahony et~al.(2005)Mahony, Hamel, and
  Pflimlin}]{mahony2005complementary}
Mahony, R., Hamel, T., and Pflimlin, J.M. (2005).
\newblock {Complementary filter design on the special orthogonal group SO (3)}.
\newblock In \emph{{Proceedings of the 44th IEEE Conference on Decision and
  Control}}, 1477--1484. IEEE.

\bibitem[{Vargas-Valencia et~al.(2016)Vargas-Valencia, Elias, Rocon,
  Bastos-Filho, and Frizera}]{vargas2016imu}
Vargas-Valencia, L.S., Elias, A., Rocon, E., Bastos-Filho, T., and Frizera, A.
  (2016).
\newblock {An IMU-to-body alignment method applied to human gait analysis}.
\newblock \emph{{Sensors}}, 16(12), 2090.

\bibitem[{{Wu} et~al.(2022){Wu}, {Shen}, {Xiao}, {Nguyen}, {Hecker}, and
  {Fitzek}}]{wu2022picaextension}
{Wu}, H., {Shen}, Y., {Xiao}, X., {Nguyen}, G.T., {Hecker}, A., and {Fitzek},
  F.H.P. (2022).
\newblock {Accelerating Industrial IoT Acoustic Data Separation with In-Network
  Computing}.
\newblock \emph{{IEEE Internet of Things Journal}}, 1--15.
\newblock \doi{10.1109/JIOT.2022.3176974}.

\bibitem[{Wu et~al.(2021)Wu, Xiang, Nguyen, Shen, and Fitzek}]{wu2021computing}
Wu, H., Xiang, Z., Nguyen, G.T., Shen, Y., and Fitzek, F.H. (2021).
\newblock {Computing meets network: Coin-aware offloading for data-intensive
  blind source separation}.
\newblock \emph{{IEEE Network}}, 35(5), 21--27.

\bibitem[{Yi et~al.(2021)Yi, Zhou, and Xu}]{yi2021transpose}
Yi, X., Zhou, Y., and Xu, F. (2021).
\newblock {TransPose: real-time 3D human translation and pose estimation with
  six inertial sensors}.
\newblock \emph{{ACM Transactions on Graphics (TOG)}}, 40(4), 1--13.

\end{thebibliography}

	\appendix
	\section{Iterative Computation for Jacobian Matrices} \label{Appendix_Jacobian}
	In this section, we derive the Jacobian matrix $\bm{H}_k(\bm{\theta})$ between the system outputs $\bm{y}_k$ in \eqref{eqn_IMU_measurement} and system states $\bm{x}_k$ in \eqref{eqn_system_state} at time $t_k, \forall k \!\in\! \mathbb{N}$.
	For notational convenience, the index $k$ representing the time stamp is neglected. 
	Then, the Jacobian matrices between the motion properties and the system states are shortened as
	\begin{align}
		&\bm{T}_{r,i} = \frac{\partial {_I\ddot{\bm{r}}^i}}{\partial \ddot{\bm{q}}} = \frac{\partial {_I\dot{\bm{r}}^i}}{\partial \dot{\bm{q}}} = \frac{\partial {_I{\bm{r}}^i}}{\partial {\bm{q}}} ~,~ 
		\dot{\bm{T}}_{r,i} = \frac{\partial {_I\dot{\bm{r}}^i}}{\partial {\bm{q}}} \nonumber
		\\
		&\dot{\bm{T}}_{r,i}^* = \frac{\partial {_I\ddot{\bm{r}}^i}}{\partial \dot{\bm{q}}} ~,~
		\ddot{\bm{T}}_{r,i}^* = \frac{\partial {_I\ddot{\bm{r}}^i}}{\partial {\bm{q}}}, \nonumber 
		\\
		&\bm{T}_{\omega,i} = \frac{\partial {_i\dot{\bm{\omega}}^i}}{\partial \ddot{\bm{q}}} = \frac{\partial {_i{\bm{\omega}}^i}}{\partial \dot{\bm{q}}} = \frac{\partial {{\bm{\varphi}}^{i,*}}}{\partial {\bm{q}}} ~,~ 
		\dot{\bm{T}}_{\omega,i} = \frac{\partial {_i{\bm{\omega}}^i}}{\partial {\bm{q}}} \nonumber
		\\
		&\dot{\bm{T}}_{\omega,i}^* = \frac{\partial {_i\dot{\bm{\omega}}^i}}{\partial \dot{\bm{q}}} ~,~
		\ddot{\bm{T}}_{\omega,i}^* = \frac{\partial {_i\dot{\bm{\omega}}^i}}{\partial {\bm{q}}},  \nonumber
	\end{align}
	where ${\bm{\varphi}}^{i,*}$ represents the integral of the ${_i{\bm{\omega}}^i}$, which is usually different as $\bm{\varphi}_i = {_i\bm{n}} q_i$.
	
	Then, we first derive the Jacobian matrix between the measurements from the $j$-th IMU attached on the $i$-th body.
	Considering that the measured translational acceleration of the IMU $_j{\bm{a}}^j$ includes the gravity and is expressed in its local coordinate, the Jacobian matrix between $\bm{y}^j$ and $\bm{x} = [\bm{q},\dot{\bm{q}},\ddot{\bm{q}}]^T$ is written as
	\begin{align} \label{eqn_SingleIMUJacobian}
		&\bm{H}^j(\bm{\theta}) =  
		\begin{bmatrix}
			\bm{R}_j^T \ddot{\bm{T}}_{r,j}
			& \bm{R}_j^T	\dot{\bm{T}}_{r,j}^* - {_j \bm{\omega} ^j_{\times}} \bm{R}_j \bm{T}_{r,j}
			& \bm{R}_j^T  \bm{T}_{r,j}\\
			\dot{\bm{T}}_{\omega,j} & \bm{T}_{\omega,j} & \bm{0}_{3 \times n}
		\end{bmatrix}, \nonumber
		\\
		&\ddot{\bm{T}}_{r,j} =  \ddot{\bm{T}}_{r,j}^* +  {_I\bm{g}_{\times}} \bm{R}_j \bm{T}_{\omega,j} - \bm{R}_j {_j \bm{\omega} ^j_{\times}} \bm{R}_j^T \dot{\bm{T}}_{r,j}.
	\end{align}
	For the IMU $j$, the Jacobian matrices are calculated through the Jacobian matrices of the attached body $i$, where the rotational part of the Jacobian matrices is computed through
	\begin{align}
		& \bm{T}_{\omega,j} = \bm{R}_{s,i,j}^T {\bm{T}_{\omega,i}}, \qquad
		\dot{\bm{T}}_{\omega,j} = \bm{R}_{s,i,j}^T {\dot{\bm{T}}_{\omega,i}}, \nonumber
		\\
		& \dot{\bm{T}}_{\omega,j}^* = \bm{R}_{s,i,j}^T \dot{\bm{T}}_{\omega,i}^* + \bm{\Lambda}_{\omega}^{i,j} \bm{T}_{\omega,i}, \nonumber
		\\
		&\ddot{\bm{T}}_{\omega,j}^* = \bm{R}_{s,i,j}^T \dot{\bm{T}}_{\omega,i}^* + \bm{\Lambda}_{\omega}^{i,j} \dot{\bm{T}}_{\omega,i}, \nonumber
	\end{align}
	in which $\bm{R}_{s,i,j} = \bm{R}(_i \bm{\varphi} ^{i,j}_s)$ and ${_{j} \bm{\omega}^{j}} = \bm{T}_{\omega,j} \dot{\bm{q}}$ and
	\begin{equation*}
		 \bm{\Lambda}_{\omega}^{i,j} = \bm{R}_{s,i,j}^T {_{i} \bm{\omega}^{i}_{\times}} - {_{j} \bm{\omega}^{j}_{\times}} \bm{R}_{s,i,j}^T .
	\end{equation*}
	And for the translational part, the Jacobian matrices are updated from the attached body $i$ as
	\begin{align}
		\bm{T}_{r,j} = & \bm{T}_{r,i} - \bm{R}_{i} {_{i} \bm{r}^{i,j}_{\times}} {\bm{T}_{\omega,i}}, \nonumber
		\\
		\dot{\bm{T}}_{r,j} = & \dot{\bm{T}}_{r,i} - \bm{R}_{i} ( {_{i} \bm{\omega}^{i}_{\times}} {_{i} \bm{r}^{i,j}_{\times}} {\bm{T}_{\omega,i}} + {_{i} \bm{r}^{i,j}_{\times}} {\dot{\bm{T}}_{\omega,i}} ), \nonumber
		\\
		\dot{\bm{T}}_{r,j}^* = & \dot{\bm{T}}_{r,i}^* -  \bm{R}_{i} ( {_{i} \bm{r}^{i,j}_{\times}} \dot{\bm{T}}_{\omega,i}^* + \bm{\Lambda}_r^{i,j} \bm{T}_{\omega,i} ) , \nonumber
		\\
		\ddot{\bm{T}}_{r,j}^* = & \ddot{\bm{T}}_{r,i}^* - \bm{R}_{i} ( {_{i} \bm{r}^{i,j}_{\times}} \ddot{\bm{T}}_{\omega,i}^* + \bm{\Lambda}_r^{i,j} \dot{\bm{T}}_{\omega,i} + {_{i} \bm{\omega}^{i}_{\times}} {_{i} \bm{\omega}^{i}_{\times}} {_{i} \bm{r}^{i,j}_{\times}} \bm{T}_{\omega,i} ), \nonumber
	\end{align}
	where
	\begin{align*}
		\bm{\Lambda}_r^{i,j} = ({_i\bm{\omega}^{i}_{\times}} {_{i} \bm{r}^{i,j}})_{\times} + 2{_{i} \bm{\omega}^{i}_{\times}} {_{i} \bm{r}^{i,j}_{\times}}.
	\end{align*}

	Therefore, the Jacobian matrix $\bm{H}(\bm{\theta})$ between the measurements from all IMU sensors $\bm{y}$ and system states $\bm{x}$ is the concatenation of $\bm{H}^j(\bm{\theta}), j = 1, \cdots, M$, which is written as
	\begin{align}
		\bm{H}(\bm{\theta}) = \begin{bmatrix}
			\bm{H}^1(\bm{\theta}) \\ \vdots \\ \bm{H}^M(\bm{\theta})
		\end{bmatrix}.
	\end{align}
	
	Furthermore, the Jacobian matrices for the $i$-th body are iteratively obtained from the Jacobian matrices of its predecessor body $i-1$.
	For notational simplicity, we consider the Jacobian matrices of body $i+1$ from body $i$, whose rotational part is iteratively updated through
	\begin{align} 
		\bm{T}_{\omega,i} = & \bm{R}_{q,i}^T {\bm{T}_{\omega,i-1}} + {_{i-1}\bm{n}} \bm{t}_{i}^T, \quad
		, \nonumber
		\\
		\dot{\bm{T}}_{\omega,i} = & \bm{R}_{q,i}^T {\dot{\bm{T}}_{\omega,i-1}} + {_{i-1} \bm{\omega}^{i-1}_{\times}} {_{i-1}\bm{n}} \bm{t}_{i}^T, \nonumber
		\\
		\dot{\bm{T}}_{\omega,i}^* = & \bm{R}_{q,i}^T \dot{\bm{T}}_{\omega,i-1}^* + \bm{\Lambda}_{\omega}^{i-1,i} \bm{T}_{\omega,i-1} + {_{i} \bm{\omega}^{i}_{\times}} {_{i-1}\bm{n}} \bm{t}_{i}^T, \nonumber
		\\
		\ddot{\bm{T}}_{\omega,i}^* = & \bm{R}_{q,i}^T \dot{\bm{T}}_{\omega,i-1}^* + \bm{\Lambda}_{\omega}^{i-1,i} \dot{\bm{T}}_{\omega,i-1} + {_{i-1}\bm{n}_{\times}} {_{i} \bm{\omega}^{i}_{\times}} {_{i-1}\bm{n}}  \bm{t}_{i}^T \dot{{q}}_{i},  \nonumber
	\end{align}
	where $\bm{t}_{i}$ is the $i$-th column of identical matrix $\bm{I}_{n}$, the angular velocity is ${_{i} \bm{\omega}^{i}} = \bm{T}_{\omega,i} \dot{\bm{q}}$ and
	\begin{align*}
		\bm{\Lambda}_{\omega}^{i-1,i} = \bm{R}_{q,i}^T {_{i-1} \bm{\omega}^{i-1}_{\times}} - {_{i} \bm{\omega}^{i}_{\times}} \bm{R}_{q,i}^T -  {_{i-1}\bm{n}_{\times}} \bm{R}_{q,i}^T \dot{{q}}_{i}.
	\end{align*}
	The iteration computation of the Jacobian matrices for the translational motion is expressed as 
	\begin{align} 
		\bm{T}_{r,i+1} = & \bm{T}_{r,i} - \bm{R}_{i} {_{i} \bm{r}^{i,i+1}_{\times}} {\bm{T}_{\omega,i}}, \nonumber
		\\
		\dot{\bm{T}}_{r,i+1} = & \dot{\bm{T}}_{r,i} - \bm{R}_{i} ( {_{i} \bm{\omega}^{i}_{\times}} {_{i} \bm{r}^{i,i+1}_{\times}} {\bm{T}_{\omega,i}} + {_{i} \bm{r}^{i,i+1}_{\times}} {\dot{\bm{T}}_{\omega,i}} ), \nonumber
		\\
		\dot{\bm{T}}_{r,i+1}^* = & \dot{\bm{T}}_{r,i}^* -  \bm{R}_{i} ( {_{i} \bm{r}^{i,i+1}_{\times}} \dot{\bm{T}}_{\omega,i}^* + \bm{\Lambda}_r^{i,i+1} \bm{T}_{\omega,i} ) , \nonumber
		\\
		\ddot{\bm{T}}_{r,i+1}^* = & \ddot{\bm{T}}_{r,i}^* - \bm{R}_{i} ( {_{i} \bm{r}^{i,i+1}_{\times}} \ddot{\bm{T}}_{\omega,i}^* + \bm{\Lambda}_r^{i,i+1} \dot{\bm{T}}_{\omega,i} \nonumber\\
		&\qquad\qquad\qquad\qquad\qquad\quad + {_{i} \bm{\omega}^{i}_{\times}} {_{i} \bm{\omega}^{i}_{\times}} {_{i} \bm{r}^{i,i+1}_{\times}} \bm{T}_{\omega,i} ), \nonumber
	\end{align}
	where
	\begin{align*}
		\bm{\Lambda}_r^{i,i+1} = ({_i\bm{\omega}^{i}_{\times}} {_{i} \bm{r}^{i,i+1}})_{\times} + 2{_{i} \bm{\omega}^{i}_{\times}} {_{i} \bm{r}^{i,i+1}_{\times}}.
	\end{align*}
	The Jacobian matrices of the pedestal (human trunk), i.e. $i = 0$, are set to $\bm{0}$, i.e. we assume that the motion of the human trunk is independent of joint motions.
	
\end{document}